\documentstyle[preprint,eqsecnum,aps]{revtex}

\def\lesssim{\ \raise.3ex\hbox{$<$}\kern-0.8em\lower.7ex\hbox{$\sim$}\ }
\def\gesim{\ \raise.3ex\hbox{$>$}\kern-0.8em\lower.7ex\hbox{$\sim$}\ }
\newcommand{\mib}{\bf}
\font\scripti=cmmi7
\font\scriptscripti=cmmi5
\def\sib#1{\setbox0 = \hbox{\scripti #1}
  \kern-.02em\copy0\kern-\wd0
  \kern.04em\box0} % script italic bold 
\def\ssib#1{\setbox0 = \hbox{\scriptscripti #1}
  \kern-.02em\copy0\kern-\wd0
  \kern.04em\box0} % scriptscript italic bold

\font\tenib=cmmib10 % italic bold for math
\skewchar\tenib='177 \skewchar\tenib='177 \skewchar\tenib='177
\def\pbold#1{\setbox0 = \hbox{$ #1 $}
  \kern-.022em\copy0\kern-\wd0
  \kern.011em\copy0\kern-\wd0
  \kern.011em\copy0\kern-\wd0
  \kern.011em\copy0\kern-\wd0
  \kern.011em\box0} % poorman's bold 

\begin{document}
\draft
%\preprint{HEP/123-qed}
\title{Anomalous Temperature Dependence of Spin Susceptibility around a Nonmagnetic Impurity in High-$T_{\rm c}$ Cuprate Superconductors}
\author{Y. Ohashi}
\address{Department of Physics, University of Toronto, Toronto, Ontario, Canada M5S 1A7\\
Institute of Physics, University of Tsukuba, Ibaraki 305, Japan}
%\date{}
\maketitle
\begin{abstract}
Local magnetic properties around a nonmagnetic impurity are investigated in high-$T_{\rm c}$ cuprate superconductors. We consider a model two-dimensional $d_{x^2-y^2}$-wave superconductor with strong antiferromagnetic (AF) spin fluctuations and calculate the spin susceptibility around the impurity. We show that the uniform susceptibility ($\chi(0)$), which is usually suppressed in spin-singlet superconductivity, is enhanced at low temperatures around the impurity when the impurity scattering is in the unitarity limit. This anomalous behavior is more remarkable when the impurity potential is extended spatially and the AF spin fluctuations are stronger. As the origin of this phenomenon, we point out the importance of (1) a low-energy bound state around the impurity and (2) a mode-mode coupling effect which connects the AF spin fluctuations with $\chi(0)$. Our results can explain the anomalous temperature dependence of the Knight shift on Li which was recently observed in YBa$_2$(Cu$_{1-x}$Li$_x$)$_3$O$_{6+y}$.
\par
\vskip3mm
\end{abstract}
\pacs{}
\narrowtext
%\newpage
%%%%%%%%%%%%%%%%%%%%%%%%%%%%%%%%%%%%%%%%%%%%%%%%%%%%%%%%%%%%%%%%%%%
%
%%%%%%%%%%%%%%%%%%%%%%%%%%%%%%%%%%%%%%%%%%%%%%%%%%%%%%%%%%%%%%%%%%%
%
\section{Introduction}
Recently, Bobroff et al. observed an anomalous behaviour of the uniform susceptibility ($\chi(0)$) in YBa$_2$(Cu$_{1-x}$Li$_x$)$_3$O$_{6+y}$:\cite{Bobroff} Usually $\chi(0)$ is suppressed in singlet superconductivity because of the singlet Cooper pairs. However, their Knight shift measurement clarified that $\chi(0)$ on Li increases with decreasing temperature in the superconducting state. In addition, this phenomenon was shown to be more remarkable in the under-doped regime than in the over-doped one. Since antiferromagnetic (AF) spin fluctuations are considered stronger in the former regime, this result implies the importance of the AF spin fluctuations in this phenomenon.
\par
In our previous papers,\cite{ohashi1,ohashi2} we have clarified two nonmagnetic impurity effects on the magnetic properties of high-$T_{\rm c}$ cuprates: 
\begin{enumerate}
\item[(A)] In $d$-wave superconductors, when the impurity scattering is in the unitarity limit, a low-energy bound state is formed around the impurity.\cite{Balatsky,Onishi} Then, at low temperatures, the low-energy AF spin fluctuations are enhanced by the density of states (DOS) originating from this bound state.\cite{ohashi1} This impurity effect can explain the enhancement of the low-energy AF spin fluctuations in Zn-doped YBa$_2$Cu$_3$O$_{6+x}$.\cite{Sidis,Fong} It also agrees with the increase of the nuclear spin-lattice relaxation rate on Cu in YBa$_2$(Cu$_{1-x}$Zn$_x$)$_4$O$_8$ far below the superconducting transition temperature ($T_{\rm c}$).\cite{Itoh1,Itoh2,Itoh3,Itoh4,Itoh5} 
\par
\item[(B)] In the normal state, when the local density of states around the Fermi level is enhanced near an impurity, the AF spin fluctuations are also enhanced locally. This enhancement affects $\chi(0)$ through a mode-coupling effect. Then $\chi(0)$ shows a Curie-Weiss-like temperature dependence around the impurity as observed in YBa$_2$(Cu$_{1-x}$M$_x$)$_3$O$_{6+y}$ (M=Zn, Li) in the normal state.\cite{Mahajan,Bobroff2} Recently, this mechanism was discussed also by Bulut.\cite{Bulut1,Bulut2}
\end{enumerate}
\par
Taking into account these effects, we can draw the following scenario as a possible mechanism of the anomalous temperature dependence of $\chi(0)$ observed by Bobroff et. al.: (1) When the impurity scattering is in the unitarity limit, the low-energy bound state appears around the impurity in $d$-wave superconductivity. (2) Then the AF spin fluctuations are enhanced around the impurity by the DOS originating from the bound state. (3) This local enhancement affects $\chi(0)$ through the mode-mode coupling effect mentioned in the impurity effect-(B). As a result, $\chi(0)$ increases with decreasing temperature around the impurity even in the superconducting state. 
\par
In this paper, we investigate the possibility of this scenario in high-$T_{\rm c}$ cuprates. For this purpose, we consider a model two-dimensional $d_{x^2-y^2}$-wave superconductor with strong AF spin fluctuations and with a nonmagnetic impurity. In this model, we calculate the static spin susceptibility around the impurity. We show that the local enhancement of $\chi(0)$ really occurs around the impurity owing to the two effects, (A) and (B), noted in the above.
\par
This paper is organized as follows: In $\S$2, we present our formulation. Nonmagnetic impurity effects on susceptibility are studied in $\S$3, which is followed by summary in $\S$4.
\vskip5mm
\section{Formulation}
\subsection{Model two-dimensional $d_{x^2-y^2}$-wave superconductor}
\vskip2mm
We consider an $N\times N$-two-dimensional square lattice with a nonmagnetic impurity at ${\mib R}_{\rm imp}=(0,0)$. We impose the periodic boundary condition in the $x$- and $y$-direction and put the lattice constant unity. The Hamiltonian is given by
\begin{eqnarray}
H=\sum_{i,j,\sigma}[t_{ij}c^\dagger_{i\sigma}c_{j\sigma}+{\rm h.c}]
-\mu\sum_{i}n_{i}
+U\sum_in_{i\uparrow}n_{i\downarrow}
-g\sum_{\langle ij\rangle}n_{i\uparrow}n_{j\downarrow}
+H_{\rm imp},
\label{eq.2.1}
\end{eqnarray}
where $c_{i\sigma}$ is the annihilation operator of an electron with $\sigma$-spin at the $i$-th site, $n_{i\sigma}=c^\dagger_{i\sigma}c_{i\sigma}$, and $n_i=\sum_\sigma n_{i\sigma}$; $U$ and $g$ represent the on-site Coulomb repulsion and the nearest-neighbor pairing interaction, respectively. In the paring term in eq. (\ref{eq.2.1}), $\langle ij\rangle$ means the summation over the nearest-neighbor sites. 
\par
In the hopping term in eq.(\ref{eq.2.1}), we put $t_{ij}=-t=-1$ for the nearest-neighbor hopping and $t_{ij}=t'=0.25$ for the next nearest-neighbor ones. (In this paper, $t=1$ is the unit of energy.) In addition, the chemical potential $\mu$ is chosen so that the number density of electrons can be $n=0.95$ at sites far away from the impurity. Then we obtain a nearly antiferromagnetically nested Fermi surface, which gives strong AF spin fluctuations when $U>0$.\cite{ohashi2}.
\par
The last term in eq.(\ref{eq.2.1}) describes the nonmagnetic impurity scattering. A recent experimental analysis clarified that the suppression of $T_{\rm c}$ by Li-doping in YBa$_2$Cu$_3$O$_{6+x}$ is almost equal to the case of Zn-doping.\cite{Bobroff2} Since Zn is considered a unitarity scatterer, this result indicates that Li should be also regarded as this type of scatterer. For this reason, we put $V_1=100$ as the strength of the impurity potential at ${\mib R}_{\rm imp}$. In addition, we assume that the impurity potential at the nearest-neighbor sites of the impurity is finite in order to examine the effect of the extended impurity potential in real space.\cite{note2} The resulting $H_{\rm imp}$ is then given by
\begin{equation}
H_{\rm imp}=V_1n_{{\sib R}_{\rm imp}}
+V_2\sum_{{\sib r}_j=(0,\pm 1),(\pm 1,0)}n_{{\sib r}_j}, 
\end{equation}
where ${\mib r}_j$ is the position vector of the $j$-th site.\par
\par
We apply the mean field approximation to the pairing term in eq. (\ref{eq.2.1}). Then this term is reduced to 
\begin{equation}
-g\sum_{\langle ij\rangle}n_{i\uparrow}n_{j\downarrow}
\to
\sum_{\langle ij\rangle}[\Delta_{ij}c^\dagger_{i\uparrow}c^\dagger_{j\downarrow}+\Delta^*_{ij}c_{j\downarrow}c_{i\uparrow}],
\label{eq.2.3}
\end{equation}
where $\Delta_{ij}\equiv -g\langle c_{j\downarrow}c_{i\uparrow}\rangle$ is the superconducting order parameter. In the present case ($t=1,t'=0.25,n=0.95$), the most stable superconducting state is the $d_{x^2-y^2}$-wave superconductivity. The order parameter then satisfies $\Delta_{ij}=\Delta_{ji}$ and $\Delta_{ii+x}=-\Delta_{i,i+y}$ in the region far away from the impurity. ($i+x(y)$ is the nearest-neighbor site of the $i$-th site in the $x(y)$-direction.) 
\par
\vskip5mm
\subsection{Spin susceptibility}
\vskip2mm
We calculate the following static spin susceptibility within the Hartree-Fock random phase approximation (HF-RPA):
\begin{equation}
\chi({\mib r}_i,{\mib r}_j)=
{\rm i}\int_0^\infty {\rm d}t {\rm e}^{-\delta t}
\langle
[\sigma_i^z(t),\sigma_i^z(0)]
\rangle,
\label{eq.2.4}
\end{equation}
where $\sigma_i^z(t)=n_{i\uparrow}(t)-n_{i\downarrow}(t)$ is the spin operator in the Heisenberg representation. In calculating the susceptibility, we extend our previous method\cite{ohashi2} to the case of the superconducting state: (1) We replace the Coulomb interaction term in eq. (\ref{eq.2.1}) by the one under HF approximation:
\begin{eqnarray}
U\sum_i n_{i\uparrow}n_{i\downarrow} \to
{U \over 2}\sum_i \langle n_i\rangle n_i
-{U \over 2}\sum_i \langle \sigma^z_i\rangle \sigma^z_i
+\sum_i[\Delta_i c^\dagger_{i\uparrow}c^\dagger_{i\downarrow}+
        \Delta^*_i c_{i\downarrow}c_{i\uparrow}],
\label{eq.2.5}
\end{eqnarray}
where we have dropped constant terms. In eq. (\ref{eq.2.5}), $\Delta_i\equiv U\langle c_{i\downarrow}c_{i\uparrow}\rangle$ is the $s$-wave superconducting order parameter which is induced around the impurity by the Coulomb interaction and the $d_{x^2-y^2}$-wave superconductivity.\cite{Shiba,Yoshioka,note} (2) Then the Hamiltonian can be diagonalized by the Bogoliubov transformation in real space: 
$(c^\dagger_{1\uparrow},c^\dagger_{2\uparrow},\cdots,c^\dagger_{N^2\uparrow},c_{1\downarrow},c_{2\downarrow},\cdots,c_{N^2\downarrow})=
(\gamma^\dagger_1,\gamma^\dagger_2,\cdots,\gamma^\dagger_{2N^2})W^\dagger$, where $W$ is a $2N^2\times 2N^2$-unitary matrix while $\gamma_i$ represents the annihilation operator of an eigen-state with an energy $E_i$. (3) The spin susceptibility under HF approximation ($\equiv\chi^0$) is calculated from
\begin{eqnarray}
\chi^0({\mib r}_i,{\mib r}_j)&=&
\sum_{m,m'}^{2N^2}
{f(E_m)-f(E_{m'}) \over E_{m'}-E_m}
[
W^*_{im}W_{im'}W^*_{jm'}W_{jm}+
W_{i+N^2m}W^*_{i+N^2m'}W_{j+N^2m'}W^*_{j+N^2m}
\nonumber \\
&+&
W^*_{im}W_{im'}W_{j+N^2m}W^*_{j+N^2m'}+
W_{i+N^2m}W^*_{i+N^2m'}W^*_{jm}W_{jm'}
].
\nonumber \\
\label{eq.2.6}
\end{eqnarray}
In eq. (\ref{eq.2.6}), $f(E)$ is the Fermi distribution function. (4) The spin susceptibility under HF-RPA is given by ${\hat \chi}=[1-(U/2){\hat \chi}^0]^{-1}{\hat \chi}^0$, where ${\hat \chi}\equiv\{\chi({\mib r}_i,{\mib r}_j)\}$ and ${\hat \chi}^0\equiv\{\chi^0({\mib r}_i,{\mib r}_j)\}$ are $N^2\times N^2$-matrices. (5) We execute the Fourier transformation in terms of the relative coordinate, ${\mib r}_j-{\mib r}_i$, in $\chi({\mib r}_i,{\mib r}_j)$:
\begin{equation}
\chi({\mib q},{\mib R})=\sum_j
{\rm e}^{2{\rm i}{\sib q}\cdot({\sib R}-{\sib r}_j)}
\chi(2{\mib R}-{\mib r}_j,{\mib r}_j),
\label{eq.2.7}
\end{equation}
where ${\mib R}\equiv({\mib r}_i+{\mib r}_j)/2$, and the summation is taken over all lattice sites. Since ${\mib R}$ is not always at a lattice site but it can be located between sites, we define the "on-site" spin susceptibility by\cite{ohashi2}
\begin{eqnarray}
{\bar \chi}({\mib q},{\mib R})=
\chi({\mib q},{\mib R})
&+&
{1 \over 2}\sum_{{\sib d}=(0,\pm 1/2),(\pm1/2,0)}
\chi({\mib q},{\mib R}+{\mib d})
\nonumber
\\
&+&
{1 \over 4}\sum_{{\sib d}=(1/2,\pm 1/2),(-1/2,\pm 1/2)}
\chi({\mib q},{\mib R}+{\mib d}).
\label{eq.2.8}
\end{eqnarray}
We use eq.(\ref{eq.2.8}) in the following analyses.
\par
In numerical calculations, we determine $\Delta_{ij}$, $\Delta_i$, $\langle n_i\rangle$, and $\langle \sigma^z_i\rangle$ self-consistently. In our previous paper,\cite{ohashi2} a large lattice ($N\le 52$) was necessary in order to avoid finite size effects in calculating $\chi$ in the normal state. On the other hand, this effect was found weak below $T_{\rm c}$, so that we put $N=30$. The $N$-dependence of $\chi$ is then very weak except for the case of $g=0.9$ in Fig. 3: In this case, weak size dependence still remains just below $T_{\rm c}$; however, it does not change our conclusions. As for $\chi$ above $T_{\rm c}$, we used the results in our previous paper.\cite{ohashi2}
\vskip5mm
\section{Results}
\vskip2mm
\subsection{Local enhancement of spin susceptibility below $T_{\rm c}$}
\vskip2mm
Figure 1 shows the temperature dependence of ${\bar \chi}({\mib q},{\mib R})$ in the case of $V_2=10$. In Fig. 1(a), ${\bar \chi}(0,{\mib R}=(1,1))$ decreases just below $T_{\rm c}=0.398$ owing to the suppression by superconductivity and then increases at lower temperatures. The increase is more remarkable than that in the case when the superconductivity is absent. (Compare ${\bar \chi}(0,{\mib R}=(1,1))$ with the result, "(1,1):normal": The latter is obtained by putting $g=0$.\cite{note1}) 
 The temperature dependence of ${\bar \chi}(0,{\mib R}=(1,1))$ qualitatively agrees with the Knight shift experiment on Li in YBa$_2$(Cu$_{1-x}$Li$_x$)$_3$O$_{6.97}$.\cite{Bobroff}
\par
On the other hand, the temperature dependence of ${\bar \chi}(0,{\mib R}=(5,5))$ is almost the same as that in the absence of the impurity (solid circles). This means that the anomalous temperature dependence of ${\bar \chi}(0,{\mib R})$ occurs in the vicinity of the impurity only. Indeed, when we examine the spatial dependence of ${\bar \chi}(0,{\mib R})$, we find that ${\bar \chi}(0,{\mib R})$ is enhanced only near the impurity site as shown in Fig. 2(a).
\par
We also obtain an anomalous behavior in the AF susceptibility. Below $T_{\rm c}$, Fig. 1(b) shows that ${\bar \chi}({\mib Q},{\mib R}=(1,1))$ (${\mib Q}=(\pi,\pi)$) is larger than that in the absence of the impurity (solid circle) and increases remarkably at low temperatures. This anomalous temperature dependence is not obtained at ${\mib R}=(5,5)$. At this site, ${\bar \chi}({\mib Q},{\mib R})$ is almost temperature-independent below $T_{\rm c}$. Namely, as in the case of the uniform susceptibility, the enhancement of the AF susceptibility occurs only in the vicinity of the impurity (Fig. 2(b)). 
\par
Figure 3 shows the effect of the strength of the pairing interaction on the anomalous increase of ${\bar \chi}(0,{\mib R}=(1,1))$. In this figure, we find that the temperature region where ${\bar \chi}(0,{\mib R})$ is suppressed by superconductivity becomes narrower as $g$ becomes smaller. In particular, at $g=0.9$, ${\bar \chi}(0,{\mib R}=(1,1))$ continues to increase with decreasing temperature below $T_{\rm c}=0.11$ as if the system is not in the superconducting state. Since the AF spin fluctuations are stronger at lower temperatures, Fig. 3 indicates that the local enhancement of the uniform susceptibility is more remarkable when the AF spin fluctuations are stronger.
\par
Experimentally,\cite{Bobroff} the Knight shift on Li decreases just below $T_{\rm c}$ and then increases at lower temperatures in YBa$_2$(Cu$_{1-x}$Li$_x$)$_3$O$_{6.97}$ ($T_{\rm c}=79.5$[K]). On the other hand, the Knight shift continues to increase below $T_{\rm c}$ in YBa$_2$(Cu$_{1-x}$Li$_x$)$_3$O$_{6.6}$ ($T_{\rm c}=41$[K]). Since the AF spin fluctuations are considered stronger in the latter, our result is consistent with this experimental tendency.
\par
\vskip3mm
\subsection{Effects of low-energy bound state and mode-mode coupling}
\vskip2mm
In order to explain the mechanism of the anomalous temperature dependence of the spin susceptibility, we briefly review the mode-mode coupling effect which was discussed in our previous paper:\cite{ohashi2} When we consider $\chi_{ij}$ up to $O(U)$ in a continuum system, we obtain\cite{ohashi2} 
\begin{eqnarray}
\chi({\mib q},{\mib R})=
\chi^0({\mib q},{\mib R})+
{U \over 2}
{\rm e}^{{{\rm i} \over 2}
[\bigtriangledown_{{\sib q}_2}\cdot\bigtriangledown_{{\sib R}_1}
-\bigtriangledown_{{\sib q}_1}\cdot\bigtriangledown_{{\sib R}_2}
]}
\chi^0({\mib q}_1,{\mib R}_1)\chi^0({\mib q}_2,{\mib R}_2),
\label{eq.3.1}
\end{eqnarray}
where we put ${\mib q}_i\to{\mib q}$ and ${\mib R}_i\to{\mib R}$ ($i=1,2)$ after the gradients are executed. When the system is homogeneous, $\bigtriangledown_{\sib R}\chi^0=0$. Then the gradient in terms of ${\mib q}$ does not work in eq. (\ref{eq.3.1}). In this case, $\chi({\mib q})$ is determined by $\chi^0$ at ${\mib q}$ only. On the other hand, when $\bigtriangledown_{\sib R}\chi^0\ne0$, $\bigtriangledown_{\sib q}\chi^0({\mib q})$ affects $\chi({\mib q})$, which means that $\chi^0({\mib q}')$ with ${\mib q}'\ne{\mib q}$ is necessary in obtaining $\chi({\mib q})$ (mode-mode coupling). 
\par
Within the neglect of the mode-mode coupling effect, the susceptibility is given by\cite{ohashi2}
\begin{equation}
\chi'({\mib q},{\mib R})=
{\chi^0({\mib q},{\mib R}) \over 1-(U/2)\chi^0({\mib q},{\mib R})}.
\label{eq.3.2}
\end{equation}
Figure 4 shows the comparison of ${\bar \chi}({\mib q},{\mib R})$ with ${\tilde \chi}({\mib q},{\mib R})$ which is given by eq. (\ref{eq.2.8}) with $\chi$ being replaced by $\chi'$. In Fig. 4(b), we find that the enhancement of the AF susceptibility at $0.2\lesssim T\le T_{\rm c}$ can be reproduced by ${\tilde \chi}$ apart from small deviation. When we calculate the local DOS which is defined by 
\begin{equation}
N(\omega,{\mib r}_i)=
\sum_{m=1}^{2N^2}|W_{im}|^2\delta(\omega-E_m),
\label{eq.3.3}
\end{equation}
we obtain large intensity around $\omega=0$ at ${\mib R}=(1,1)$ originating from the low-energy bound state as shown in Fig. 5.\cite{Balatsky,Onishi} Since $\chi'({\mib q},{\mib R})$ is affected by the local DOS at ${\mib R}$, the low-energy bound state is considered the origin of the local enhancement of the AF susceptibility.
\par
On the other hand, the local enhancement of the uniform susceptibility cannot be explained within the neglect of the mode-mode coupling (Fig. 4(a)). Thus, as in the case of the normal state,\cite{ohashi2} the mode-mode coupling is crucial in the case of the uniform susceptibility.
\par
Roughly speaking, when spin fluctuations are enhanced locally {\it in real space}, this phenomenon is described as the enhancement of the spin susceptibility at {\it various ${\mib q}$ in momentum space}.  In the present case, although mainly the AF spin susceptibility is enhanced by the effect of the modification of the local DOS around the impurity, the uniform susceptibility is also enhanced because of the fact that the enhancement of the spin fluctuations is {\it localized in real space}.
\par
Figure 6 shows the temperature dependence of the spin susceptibility in the case of the $\delta$-functional impurity potential ($V_2=0$). In this case, although both ${\bar \chi}(0,{\mib R})$ and ${\bar \chi}({\mib Q},{\mib R})$ at ${\mib R}=(0,0)$ are suppressed above $T_{\rm c}$ compared with those at ${\mib R}=(5,5)$, they increase far below $T_{\rm c}$ as in the case of $V_2=10$. At low temperatures, they are enhanced locally around the impurity (Fig. 7).
\par
In the normal state, the spatial dependence of the impurity potential is crucial in obtaining the Curie-Weiss-like temperature dependence of $\chi(0)$ around the impurity.\cite{ohashi2,Bulut1,Bulut2} For example, the anomalous temperature dependence is not obtained when $V_2=0$: In this case, the local enhancement of DOS around the Fermi level does not occur in contrast to the case of $V_2\ne 0$. Then the AF spin fluctuations are not enhanced around the impurity, so that the anomaly in $\chi(0)$ is also absent. On the other hand, in the $d_{x^2-y^2}$-wave superconducting state, since the low-energy bound state exists around a unitarity scatterer even in the case of $V_2=0$, the local DOS has large intensity around $\omega=0$ near the impurity (inset in Fig.6). Then the AF spin fluctuations become stronger using this large intensity of the local DOS, so that the uniform susceptibility is also enhanced by the mode-mode coupling effect. 
\par
Comparing Figs. 1 and 2 ($V_2=10$) with Figs. 6 and 7 ($V_2=0$), we find that the local enhancement of the spin susceptibility is more remarkable in the former case. Namely, detailed spatial dependence of the impurity potential is crucial in considering the local enhancement of the spin susceptibility {\it quantitatively}. 
\vskip3mm
\subsection{Local moment}
\vskip2mm
In Fig. 1, the AF phase transition occurs at $T=T_{\rm M}\simeq 0.1$. Below $T_{\rm M}$, ${\bar \chi}({\mib Q},{\mib R})$ at ${\mib R}=(1,1)$ decreases with decreasing temperature, and the spontaneous magnetic moment ($\langle\sigma_z\rangle$) grows at this site as shown in Fig. 8. This antiferromagnetism is inhomogeneous and is localized around the impurity (Fig. 9); the induced moment is largest at $(\pm 1,\pm 1)$. Because of this inhomogeneity, we obtain a finite total magnetic moment when we sum up $\langle \sigma_z\rangle$ spatially.
\par
Since the present calculation is based on the mean field approximation, this local magnetic instability itself is an artifact. We should regard $T_{\rm M}$ as a characteristic temperature at which the local moment is formed around the impurity. However, when the impurity concentration is finite, it might be possible that the local antiferromagnetism around each impurity connects to each other three-dimensionally and then the magnetism is stabilized. Physical properties below $T_{\rm M}$ and the possibility of the stabilization of this antiferromagnetism remain as our future problems.    
\vskip5mm
\section{Summary}
\vskip3mm
In this paper, we investigated the uniform susceptibility and the AF one around the nonmagnetic impurity in $d_{x^2-y^2}$-wave superconductivity. We calculated the spin susceptibility in real space taking into account the effects of the extended impurity potential, $d_{x^2-y^2}$-wave superconductivity and the Coulomb interaction. 
\par
When the impurity scattering is in the unitarity limit, the low-energy bound state appears around the impurity, which increases the intensity of the local DOS around $\omega=0$. This increase enhances the AF spin fluctuations around the impurity at low temperatures. 
This phenomenon affects $\chi(0)$ thought the mode-mode coupling effect, so that $\chi(0)$ also increases at low temperatures. In addition, the enhancement of $\chi(0)$ is more remarkable when the AF spin fluctuations are stronger. In particular, when the AF spin fluctuations are strong to some extent, $\chi(0)$ near the impurity continues to increase below $T_{\rm c}$ without being suppressed by superconductivity. These results can explain the recent Knight shift experiment in YBa$_2$(Cu$_{1-x}$Li$_x$)$_3$O$_{6+y}$.\cite{Bobroff}
\par
With regard to the effect of the extended impurity potential in real space, we showed that, apart from quantitative difference, the local enhancement of $\chi(0)$ is also obtained when the impurity potential is finite only at the impurity site. This result is in contrast to that in the normal state; above $T_{\rm c}$, the local enhancement of $\chi(0)$ is not obtained as far as we use the $\delta$-functional impurity potential. The difference originates from the existence of the low-energy bound state in the case of the $d_{x^2-y^2}$-superconducting state; it enhances the AF spin fluctuations, which increases $\chi(0)$ locally.
\par
We also showed within the mean field theory that the local antiferromagnetism may appear around the impurity below $T_{\rm M}$. Although this antiferromagnetism itself is an artifact of the present mean field approximation, we can still expect a formation of a local moment around $T_{\rm M}$. Since the present mean field theory is inadequate to examined the region below $T_{\rm M}$, a more sophisticated theory is necessary in order to clarify magnetic properties below $T_{\rm M}$. This problem remains as our future problem.
%
%
%%%%%%%%%%%%%%%%%%%%%%%%%%%%%%%%%%%%%%%%%%%%%%%%%%%%%%%%%%%%%%%%%%%
%
\acknowledgements
I would like to thank Mr. M. Terao for sending the list of numerical data from Japan to Canada. Thanks are due to Dr. Y. Itoh for valuable discussions on Zn-doped high-$T_{\rm c}$ cuprates. I also thank Professor A. Griffin for his hospitality during my stay in University of Toronto. 

\newpage
%%%%%%%%%%%%%%%%%%%
%\reference*

%%%%%%%%%%%%%%%%%%%%%%%%%%%%%%%%%%%%%%%%%%%%%%%%%%%%%%%%%%%%%%%%%%%%%%%%%%%%%%%
\newpage
\centerline{\bf Figure Captions}
\begin{enumerate}
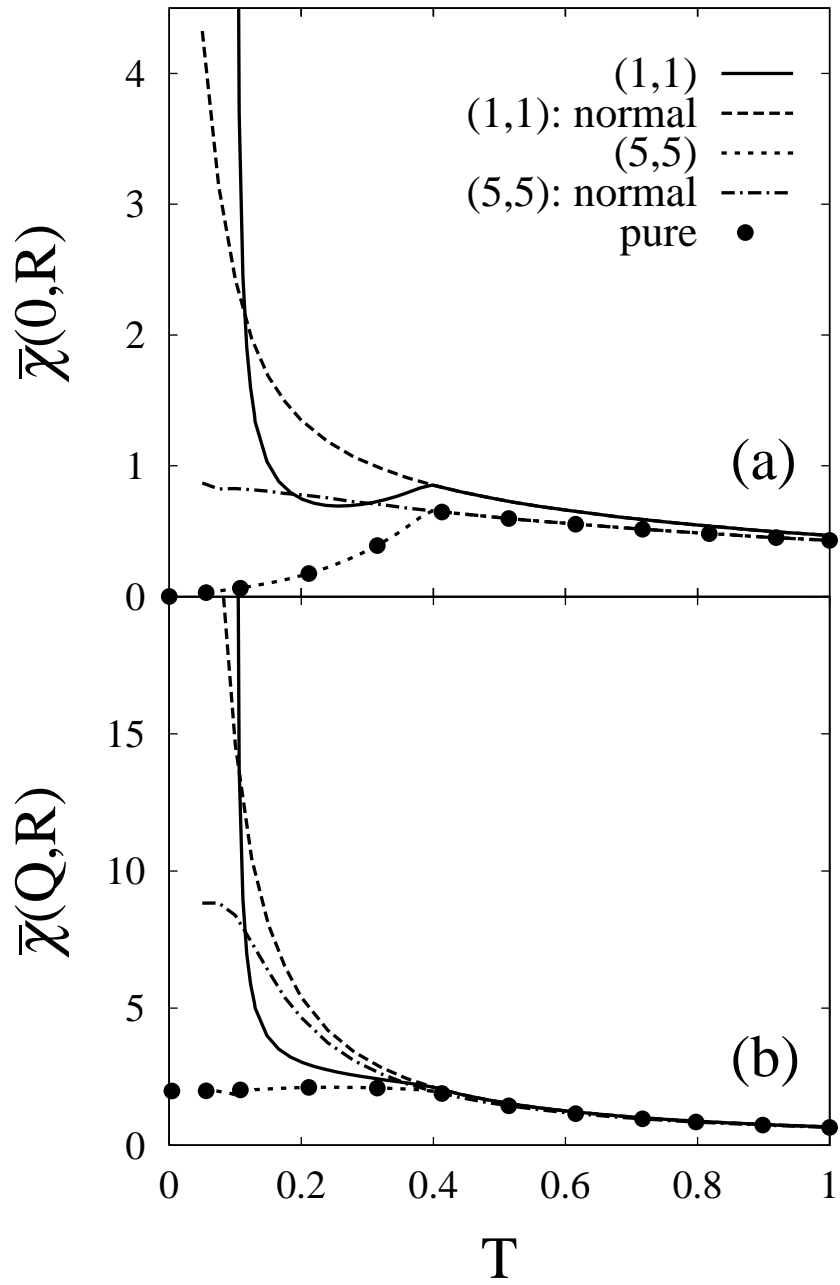
\item[Fig.1:] Temperature dependence of spin susceptibility in the case of $V_2=10$. (a) uniform susceptibility, ${\bar \chi}(0,{\mib R})$. (b) AF susceptibility, ${\bar \chi}({\mib Q},{\mib R})$. Here, for example, $(1,1)$ is the result for ${\mib R}=(1,1)$. In this figure and in Figs. 3, 4, and 6, 'normal' means that the system is in the normal state ($g=0$), while 'pure' is the case in the absence of the impurity. We put $U=2$, $g=2$ and $V_1=100$; in this case, we obtain $T_{\rm c}=0.398$. These values are used also in the following figures except for Fig. 3. At $T=T_{\rm M}\simeq 0.1$, the AF phase transition occurs. (${\bar \chi}({\mib q},{\mib R})$ below $T_{\rm M}$ is shown in Fig. 8.) 
\par
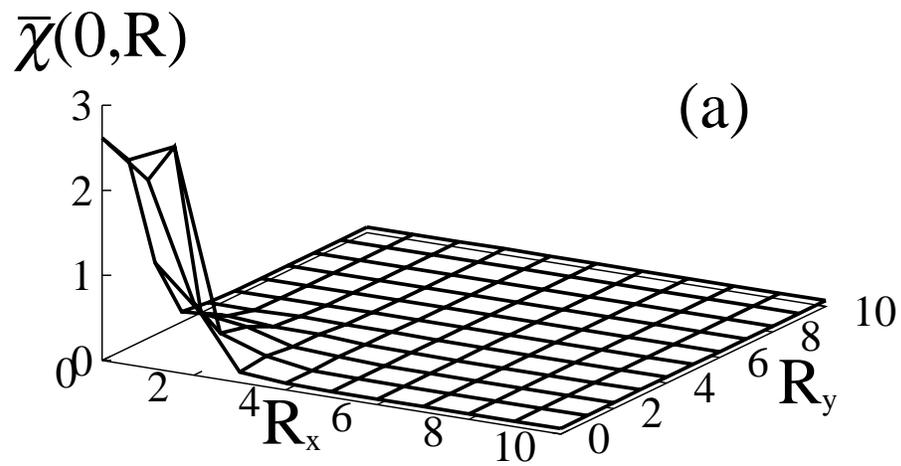
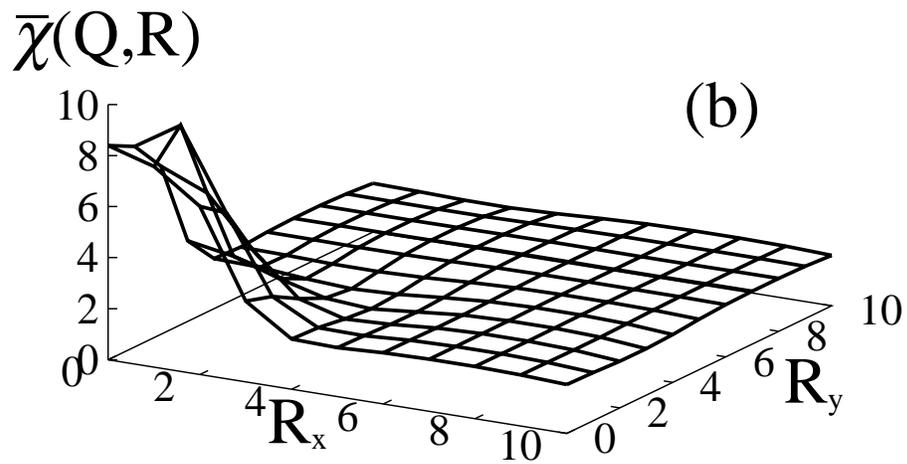
\item[Fig.2:] Spatial dependence of ${\bar \chi}({\mib q},{\mib R})$ at $T=0.1116$ in the case of $V_2=10$. 
\par
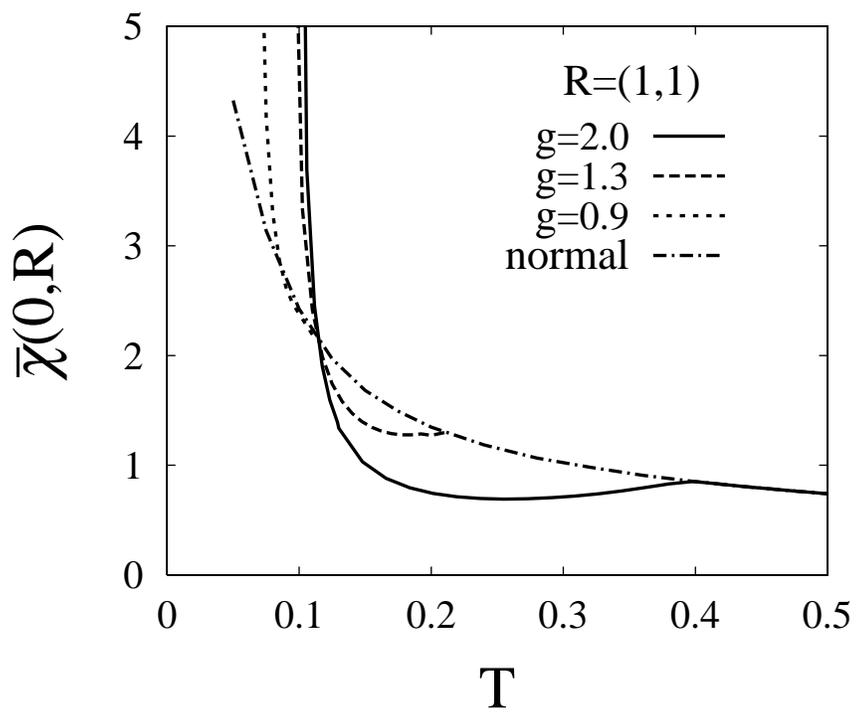
\item[Fig.3:] $g$-dependence of the uniform susceptibility at ${\mib R}=(1,1)$. We put $V_2=10$ and $U=2$. Transition temperatures are $T_{\rm c}=0.398$ ($g=2$), $T_{\rm c}=0.21$ ($g=1.3$), and $T_{\rm c}=0.11$ ($g=0.9$).
\par
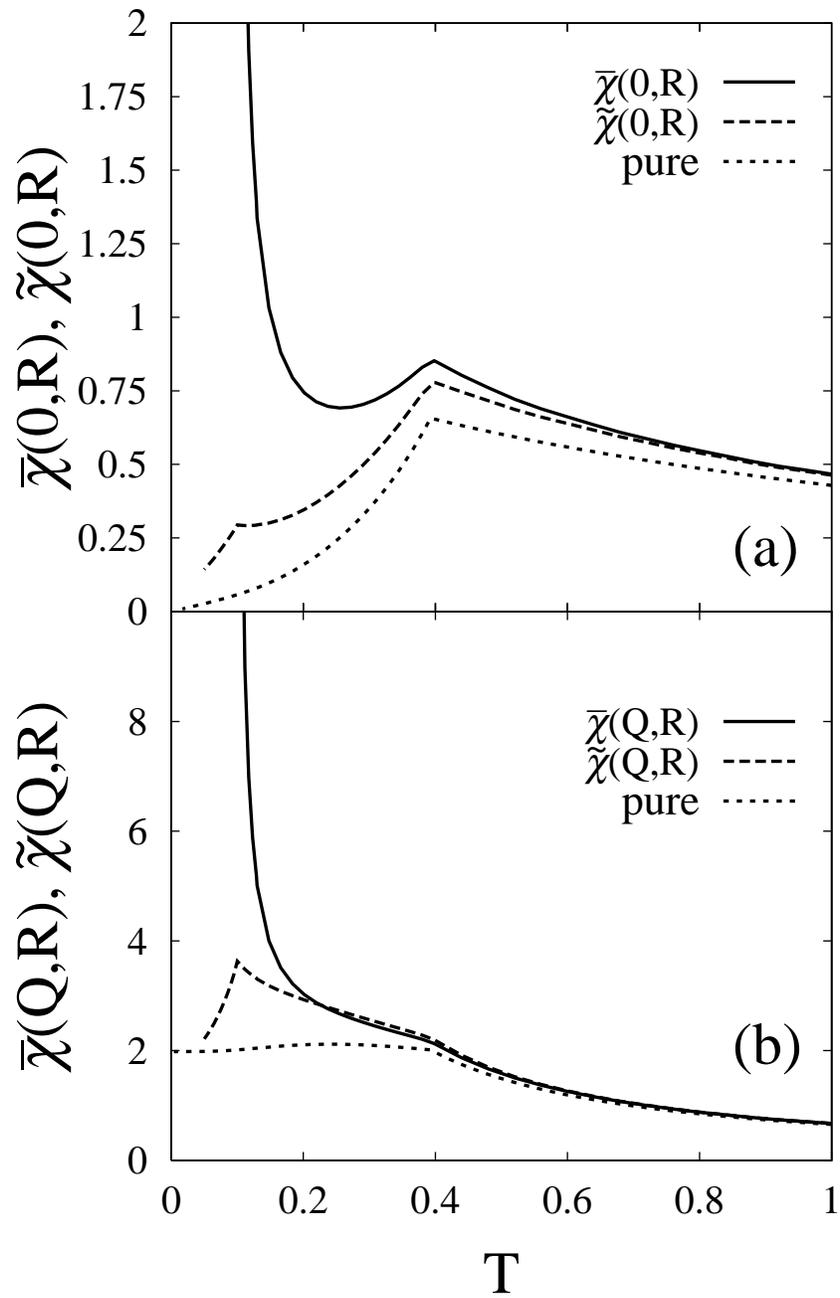
\item[Fig.4:] Comparison of ${\bar \chi}({\mib q},{\mib R})$ with ${\tilde \chi}({\mib q},{\mib R})$ at ${\mib R}=(1,1)$. The cusp in ${\tilde \chi}({\mib q},{\mib R})$ at $T\simeq 0.1$ is due to the AF phase transition at $T_{\rm M}$.
\par
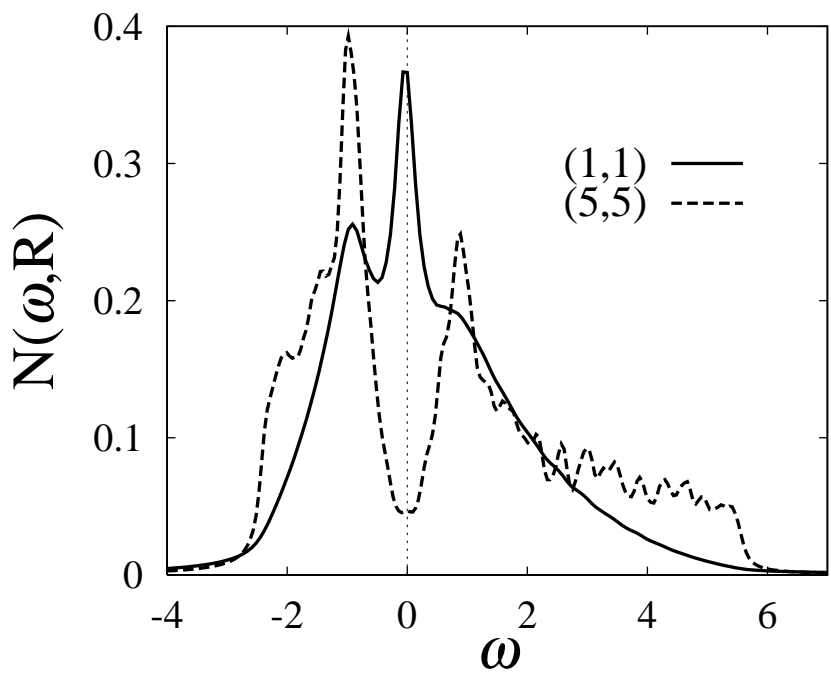
\item[Fig.5:] Local DOS at $T=0.15$. In calculating the DOS, we put $N=50$ and added small imaginary part $\Gamma=0.2$ to eigen-energies.
\par
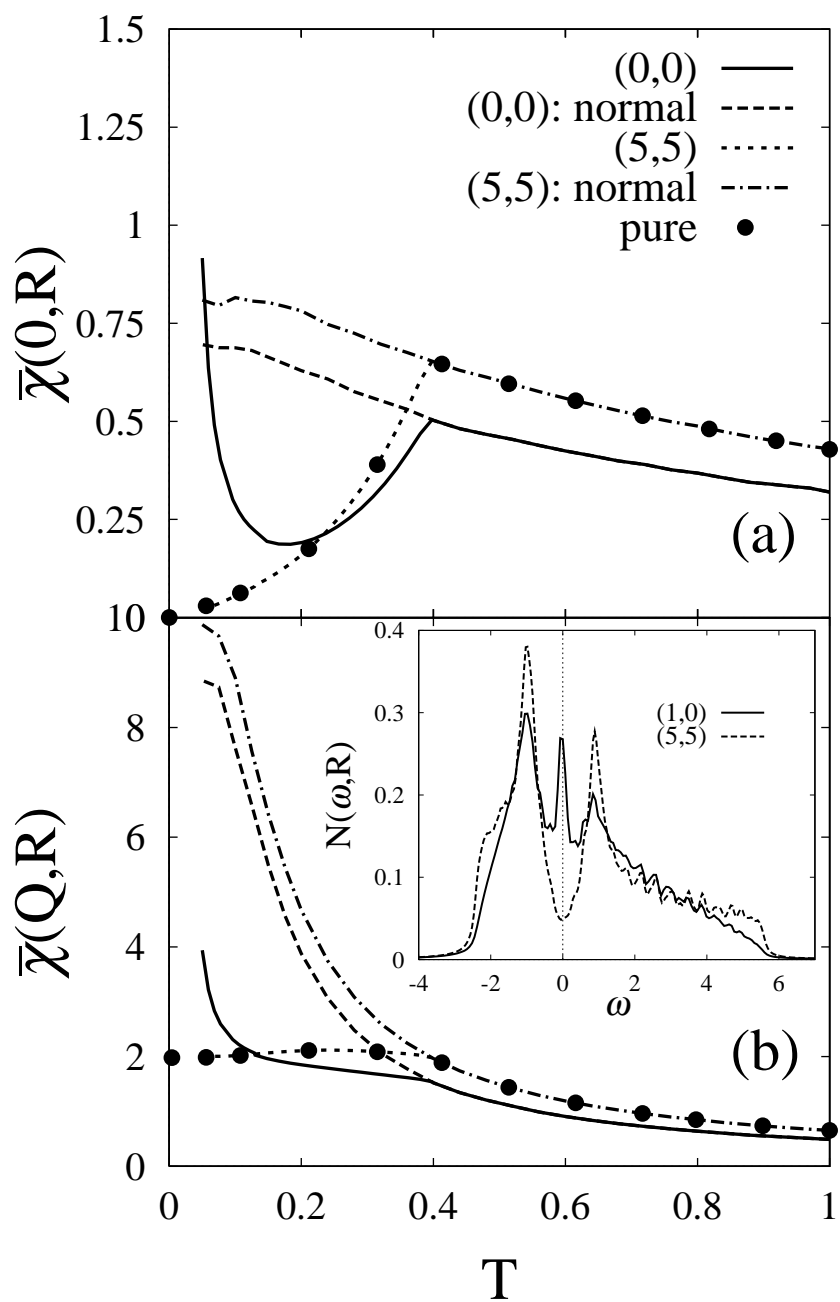
\item[Fig.6:] Temperature dependence of spin susceptibility around the impurity in the case of $V_2=0$. The inset shows the local DOS at $T=0.15$.
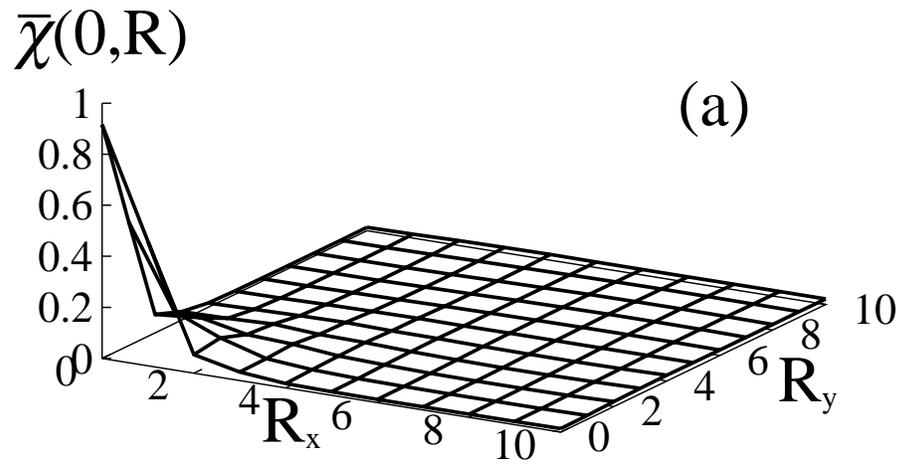
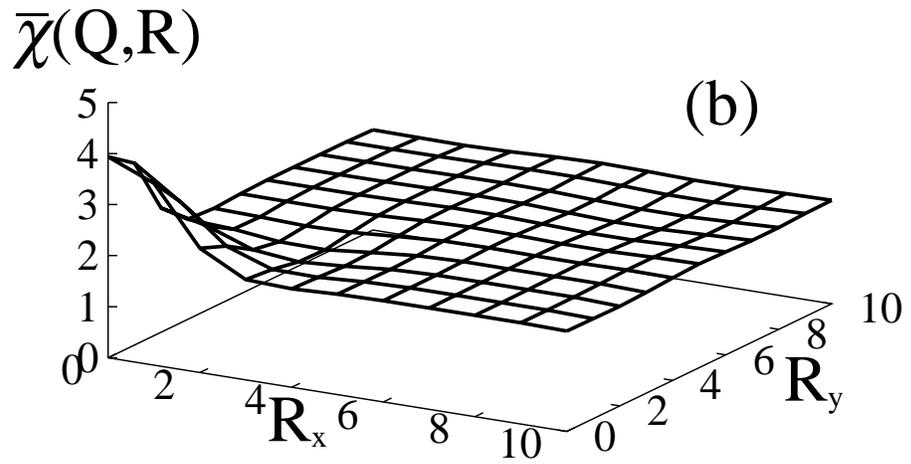
\item[Fig.7:] Spatial dependence of ${\bar \chi}({\mib q},{\mib R})$ at $T=0.05$ in the case of $V_2=0$. 

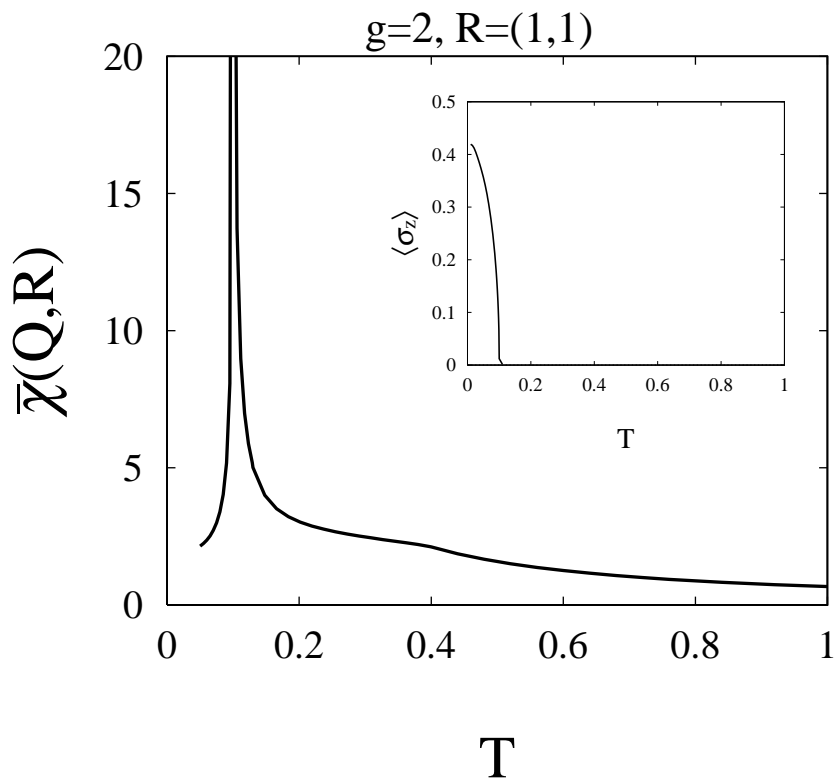
\item[Fig.8:] Temperature dependence of ${\bar \chi}({\mib Q},{\mib R})$ above and below $T_{\rm M}$ at ${\mib R}=(1,1)$ ($V_2=10$). The inset shows the spontaneous magnetic moment $\langle\sigma_z\rangle$ at ${\mib R}=(1,1)$.
\par
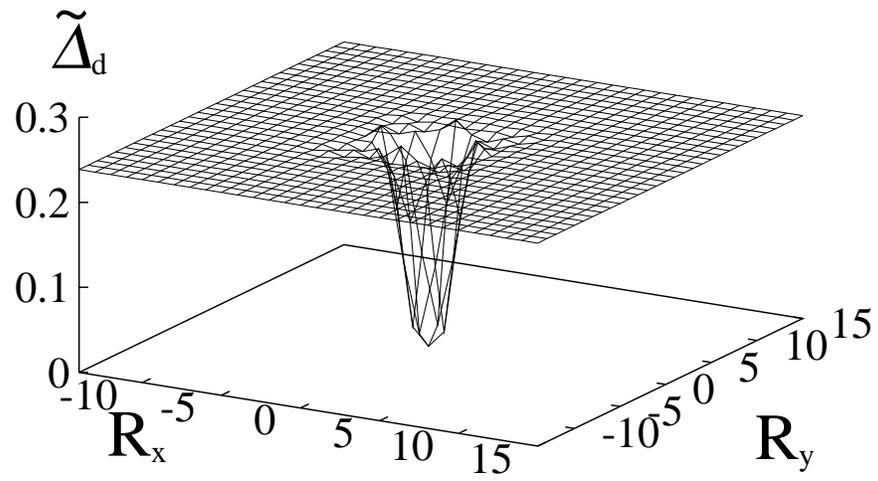
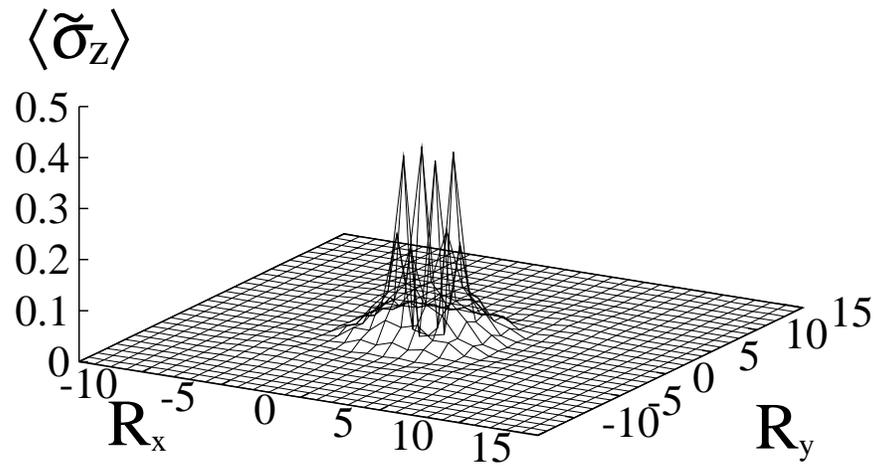
\item[Fig.9:] Spatial dependence of (a) the $d_{x^2-y^2}$-wave order parameter and (b) the spontaneous magnetic moment. We put $V_2=10$ and $T=0.05$. In this figure, 
${\tilde \Delta}_d({\mib r}_i)\equiv
{1 \over 4}
\sum_{{\sib d}=(\pm 1,0),(0,\pm 1)}
\Delta_{{\sib r}_i,{\sib r}_i+{\sib d}}\cdot(-1)^{d_y}$ 
and
${\tilde \sigma}^z({\mib r}_i)\equiv 
\sigma^z_i\cdot{\rm e}^{\pi{\rm i}[x_i+y_i]}$ (${\mib r}_i=(x_i,y_i)$).
\end{enumerate}

\begin{thebibliography}{99}
\bibitem{Bobroff} J. Bobroff, H. Alloul, W. A. MacFarlane, P. Mendels, N. Blanchard, G. Collin and J. -F. Marucco: Phys. Rev. Lett. {\bf 86} (2001) 4116.
\bibitem{ohashi1} Y. Ohashi: J. Phys. Soc. Jpn. {\bf 69} (2000) 2977.
\bibitem{ohashi2} Y. Ohashi: J. Phys. Soc. Jpn. {\bf 70} (2001) 2054.
\bibitem{Balatsky} A. V. Balatsky, M. I. Salkola and A. Rosengren: Phys. Rev. B {\bf 51} (1995) 15547.
\bibitem{Onishi} Y. Onishi, Y. Ohashi, Y. Singaki and K. Miyake: J. Phys. Soc. Jpn. {\bf 65} (1996) 675.
\bibitem{Sidis} Y. Sidis, P. Bourges, B. Hennion, L. P. Regnault, R. Villeneuve, G. Collin, F. -F. Marucco: Phys. Rev. B {\bf 53} (1996) 6811.
\bibitem{Fong} H. F. Fong, P. Bourges, Y. Sidis, L. P. Regnault, J. Bossy, A. Ivanov, D. L. Millius, I. A. Aksay and B. Keimer: Phys. Rev. Lett. {\bf 82} (1999) 1939.
\bibitem{Itoh1} Y. Itoh, T. Machi and M. Koshizuka: {\it Advances in Superconductivity, XII} (Springer Verlag, Tokyo, 2000) p. 284.
\bibitem{Itoh2} Y. Itoh, T. Machi, N. Watanabe, S. Adachi and N. Koshizuka: Physica C {\bf 357-360} (2001) 69.
\bibitem{Itoh3} Y. Itoh, T. Machi, N. Watanabe, S. Adachi and N. Koshizuka: J. Phys. Soc. Jpn. {\bf 70} (2001) 1881.
\bibitem{Itoh4} Y. Itoh, T. Machi, N. Watanabe and N. Koshizuka: J. Phys. Soc. Jpn. {\bf 70} (2001) 644.
\bibitem{Itoh5} Y. Itoh, T. Machi, N. Watanabe,S. Adachi, C. Kasai and N. Koshizuka: cond-mat/0110548.
\bibitem{Julien} M.-H. Julien, T. Feh\'er, M. Horvati\'c, C. Berthier, O. N. Bakharev. P. S\'egansan, G. Collin and J.-F. Marucco: Phys. Rev. Lett. {\bf 84} (2000) 3422.
\bibitem{Mahajan}A. V. Mahajan, H. Alloul, G. Collin and J.-F. Marucco: Phys. Rev. Lett. {\bf 72} (1994) 3100; Eur. Phys. J. B {\bf 13} (2000) 457.
\bibitem{Bobroff2} J. Bobroff. W. A. MacFarlane, H. Alloul, P. Mendels, N. Blanchard, G, Collin and J.-F. Marucco: Phys. Rev. Lett. {\bf 83} (1999) 4381.
\bibitem{Bulut1} N. Bulut: Phys. Rev. B {\bf 61} (2000) 9051.
\bibitem{Bulut2} N. Bulut: cond-mat/0108173.
\bibitem{note2} References 3,13, and 14 point out the importance of the spatial variation of the impurity potential in explaining the local enhancement of $\chi(0)$ observed in the normal state of YBa$_2$(Cu$_{1-x}$M$_x$)$_3$O$_{6+y}$ (M=Zn, Li). We also mention that an analysis of the residual resistivity in YBa$_2$(Cu$_{1-x}$Zn$_x$)O$_{6.9}$ indicates that the impurity potential of Zn is extended spatially: T. Xiang and J. M. Wheatley: Phys. Rev. B {\bf 51} (1995) 11721. \bibitem{Shiba} H. Shiba: Prog. Theor. Phys. {\bf 50} (1973) 50.
\bibitem{Yoshioka} T. Yoshioka and Y. Ohashi: J. Phys. Soc. Jpn. {\bf 69} (2000) 1812.
\bibitem{note} Actually, this induced order parameter is negligibly small compared with the magnitude of $\Delta_{ij}$ within our calculation. 
\bibitem{note1} The Curie-Weiss-like temperature dependence of the uniform susceptibility in the normal state ('(1,1):normal' in Fig. 1(a)) is due to the impurity effect-(B) explained in the introduction. As for the detail of this effect, see Ref. 3.
\end{thebibliography}
\end{document}